\documentclass[twocolumn,showpacs,preprintnumbers,amsmath,amssymb]{revtex4}
\usepackage{amsmath}
\usepackage{amssymb}
\usepackage{amsthm}
\usepackage{epsfig}
\usepackage{subfigure}
\usepackage{graphics}
\usepackage{float}
\usepackage{latexsym}
\usepackage{rotating}

\newcommand{\bra}[1]%
   {\ensuremath{\langle \, #1 \, |}}
\newcommand{\ket}[1]%
   {\ensuremath{| \, #1 \, \rangle}}
\newcommand{\braket}[2]%
   {\ensuremath{\langle \, #1 \, | \, #2 \, \rangle}} 
\newcommand{\expect}[1]%
   {\ensuremath{\langle \, #1 \,  \rangle}}
\newcommand{\rd}{\mathrm{d}}

\newcommand{\beq}{\begin{equation}}
\newcommand{\eeq}{\end{equation}}
\newcommand{\ba}{\begin{array}}
\newcommand{\ea}{\end{array}}
\newcommand{\beqa}{\begin{eqnarray}}
\newcommand{\eeqa}{\end{eqnarray}}

\usepackage{graphicx}
\usepackage{dcolumn}
\usepackage{bm}
\usepackage{dpscolor}
\usepackage{epsfig}

\begin{document}


\title{
Measurement of the Partial Cross Sections $\sigma_{\text{TT}}$,
  $\sigma_{\text{LT}}$, and $(\sigma_{\text{T}} + \varepsilon
  \sigma_{\text{L}})$ of the $p(e,e' \pi^{+})n$ Reaction in the
  $\Delta(1232)$ Resonance. }

\author{J.M.~Kirkpatrick$^1$,  N.F.~Sparveris$^2$\footnote{current address: Massachusetts Institute of Technology},
I.~Nakagawa$^3$, A.M.~Bernstein$^3$\footnote{corresponding author,
e-mail address: bernstein@mit.edu}, R.~Alarcon$^4$, W.~Bertozzi$^3$,
T.~Botto$^3$, P.~Bourgeois$^5$, J.~Calarco$^1$, F.~Casagrande$^3$,
M.O.~Distler$^6$, K.~Dow$^3$, M.~Farkondeh$^3$,
S.~Georgakopoulos$^2$, S.~Gilad$^3$, R.~Hicks$^5$, A.~Holtrop$^1$,
A.~Hotta$^5$, X.~Jiang$^5$, A.~Karabarbounis$^2$, S.~Kowalski$^3$,
R.~Milner$^3$, R.~Miskimen$^5$, C.N.~Papanicolas$^2$,
A.J.~Sarty$^7$, Y.~Sato$^8$, S.~\v Sirca$^3$, J.~Shaw$^5$,
E.~Six$^4$, S.~Stave$^3$\footnote{current address: Triangle
Universities Nuclear Laboratory, Duke University, Durham, North
Carolina 27708, USA}, E.~Stiliaris$^2$, T.~Tamae$^8$,
G.~Tsentalivich$^3$, C.~Tschalaer$^3$, W.~Turchinetz$^3$,
Z-L.~Zhou$^3$ and T.~Zwart$^3$}

\affiliation{$^1$Department of Physics, University of New Hampshire,
   Durham, New Hampshire 03824, USA}

\affiliation{$^2$Institute of Accelerating Systems and Applications
and Department of Physics, University of Athens, Athens, Greece}

\affiliation{$^3$Department of Physics, Laboratory for Nuclear
   Science, and Bates Linear Accelerator Center, Massachusetts Institute
   of Technology, Cambridge, Massachusetts 02139, USA}

\affiliation{$^4$Department of Physics and Astronomy, Arizona
   State University, Tempie, Arizona 85287, USA}

\affiliation{$^5$Department of Physics, University of
   Massachusetts, Amherst, Massachusetts 01003, USA}

\affiliation{$^6$Institut fur Kernphysik, Universit\"at Mainz,
Mainz, Germany}

\affiliation{$^7$Department of Astronomy and Physics, St. Mary's
   University, Halifax, Nova Scotia, Canada}

\affiliation{$^8$Laboratory for Nuclear Science, Tohoku
   University, Mikamine, Taihaku-ku, Sendai 982-0826, Japan}

\date{\today}

\begin{abstract}

  We report new precision $p(e,e' \pi^{+})n$ measurements in the
  $\Delta(1232)$ resonance at $Q^2 = 0.127$(GeV/c)$^2$  obtained
  at the MIT-Bates Out-Of-Plane scattering facility. These are the
  lowest, but non-zero, $Q^2$  measurements in the $\pi^+$ channel.
  The data offer new tests of the  theoretical calculations, particularly
  of the background amplitude contributions. The chiral effective field theory
  and Sato-Lee model calculations are not in agreement with this experiment.

\end{abstract}


\maketitle

Hadrons are composite systems with non-spherical quark-gluon and
meson-nucleon interactions, so there is no reason to expect that
they will be spherical \cite{Ru75,is82}. For the past few decades
there has been  extensive work to measure and quantify the deviation
from spherical symmetry based on the $ \gamma^{*} N \rightarrow
\Delta$ reaction. For recent reviews and references to the
literature see \cite{workshop,rev,pvy}.


The spectroscopic quadrupole moment provides the most reliable and
interpretable measurement of the presence of non-spherical
components in the wavefunction. For the proton it vanishes identically because of its spin 1/2 nature.
Instead, the magnitude of the non-spherical components are measured
by the resonant   electric quadrupole (E2) and Coulomb quadrupole
(C2) amplitudes $E^{3/2}_{1+}, S^{3/2}_{1+}$(for the notation see
\cite{notation,multi}) in
the predominantly magnetic dipole (M1) $M^{3/2}_{1+}$ $\gamma^*
N\rightarrow \Delta$ transition \cite{workshop,rev,pvy}.
Nonvanishing resonant quadrupole amplitudes will signify
that either the proton or the $\Delta^{+}(1232)$ or more likely both
are characterized by non-spherical components in their wavefunctions.

In the constituent-quark picture of hadrons, these non-spherical
amplitudes are a consequence of the non-central color-hyperfine
interaction among quarks \cite{Ru75,is82}. However it has been shown
that  this mechanism only provides a small fraction of the observed
quadrupole signal at low $Q^{2}$ \cite{workshop,rev,pvy}. At long
ranges the dominant contribution originates in the spontaneous
breaking of chiral symmetry which leads to a spherically asymmetric
virtual pion cloud \cite{rev}. With this in mind our group has
foucused on the $\gamma^{*} p \rightarrow \Delta$ reaction at low
$Q^{2} (\leq 0.2~(GeV/c)^{2})$
\cite{spaprc,kun00,spaprl,stave,spaplb,staveprc,spavcs} with precise
measurements at Bates and Mainz. Other experiments at low $Q^{2}$
have also been performed at Mainz and Brookhaven
\cite{pho1,pho2,pos01,bart}. At Jefferson Lab a series of
experiments have been performed primarily at higher $Q^{2}$ and
mostly in the $\pi^{0}$ channel
\cite{frol,joo,kelly,joo1,joo2,smith,ungaro,clas-pip,park}.

With the existence of non-spherical components in the nucleon
wavefunction well established, recent investigations have focused on
testing the reaction calculations and reducing the errors in
extracting the resonant multipoles from the data. To do this  we
have explored all three reaction channels associated with the
$\gamma^* N\rightarrow \Delta$ transition: H$(e,e^\prime p)\pi^0$,
H$(e,e^\prime \pi^+)n$ and H$(e,e^\prime p)\gamma$. So far at the
kinematics of this work, at $Q^2 = 0.127~$(GeV/c)$^2$, and in the
low $Q^{2}$ region, only the H$(e,e^\prime p)\pi^0$ channel has been
extensively explored. The one exception is the CLAS experiment for
$0.25 \leq Q^{2} \leq 0.65~(GeV/c)^{2}$ \cite{clas-pip}.

The exploitation of the other
two reaction channels offers us the unique opportunity to explore
the physics of interest through different, complementary, reaction
channels that can provide additional information with respect to
both the resonant and the background amplitude contributions.
The background contributions
need to be reasonably accurately known for the determination of the weak quadrupole resonant amplitudes. These play a more significant role in the $\pi^{0}$ channel off resonance, or in the $\pi^{+}$ channel. They also are dominant in the $TL^{'}$ (fifth structure function) observable. This gives three different ways to test these contributions.

Exploring the
H$(e,e^\prime \pi^+)n$ reaction channel (complementary to
H$(e,e^\prime p)\pi^0$) offers the necessary information in order to
separate the two isospin terms ($I$=1/2 and $I$=3/2) that both
contribute with different weighting in the two channels
\begin{align}
  A (\gamma p \rightarrow n \pi^+)
      &= \sqrt{2}\left (A_p^{1/2}
                        - \frac{1}{3} A^{(3/2)} \right ) \\
  A (\gamma p \rightarrow p \pi^0)
      &= A_p^{1/2} + \frac{2}{3} A^{(3/2)}
\end{align}
where A is any multipole operator.

The cross section of the H$(e,e' \pi^+)n$ reaction is sensitive to
four independent partial cross sections:
\begin{align}
  \begin{split}
   \frac{\rd \sigma}{\rd \Omega_{e} \,\rd \Omega_\pi^{*} \,\rd \omega}
    =& \, \Gamma \left ( \sigma_{T} + \varepsilon \sigma_{L} + \varepsilon
    \sigma_{TT} \cos{2 \phi_{\pi q}} \right . \\
          & \quad  \left . + v_{LT} \,
        \sigma_{LT} \cos{\phi_{\pi q}} \right ) \, ,
  \end{split}
\end{align}
were $\varepsilon$  is the transverse polarization of the virtual
photon, $\Gamma$ is the virtual photon flux, $\phi_{\pi q}$ is the
pion azimuthal angle with respect to the momentum transfer
direction, and $v_{LT} = \sqrt{2 \varepsilon (1 + \varepsilon)}$.
The $\sigma_T$ and $\sigma_L$ terms can be combined into a single
one, $\sigma_{0} \equiv \sigma_{T} + \varepsilon \sigma_{L}$.

The E2 and C2 amplitudes manifest themselves mostly through the
interference with the dominant dipole (M1) amplitude. The
interference of the C2 amplitude with the M1 leads to the
longitudinal-transverse (LT) response while the interference of the
E2 amplitude with the M1 leads to the transverse-transverse (TT)
response. The $\sigma_0$ partial cross section is dominated by the
M1 multipole.

\begin{figure}[h]
\centerline{\psfig{figure=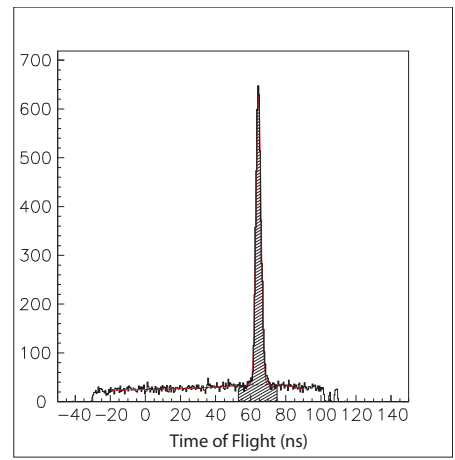,height=6.0cm}}
\smallskip
\caption{The coincidence time of flight spectrum.} \label{fig:tof}
\end{figure}

\begin{figure}[h]
\centerline{\psfig{figure=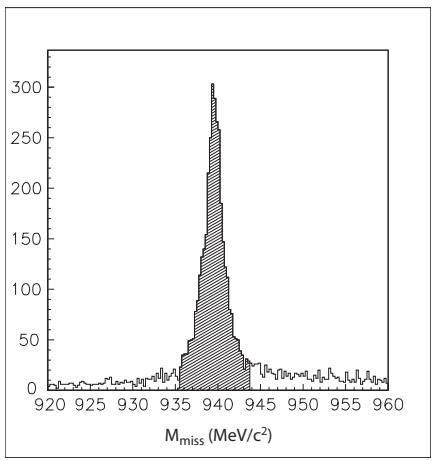,height=6.0cm}}
\smallskip
\caption{Typical missing mass distribution for the reconstructed
neutron.} \label{fig:mmiss}
\end{figure}

The H$(e,e' \pi^+)n$ measurements were performed using the
out-of-plane spectrometer (OOPS) system \cite{oops} of the MIT-Bates
Laboratory. Electrons were detected with the OHIPS spectrometer
\cite{ohips} which employed two vertical drift chambers for the
track reconstruction and three scintillator detectors for timing
information. A Cherenkov detector and two layers of 18 Pb-glass
detectors were used for particle identification. Pions were detected
with the OOPS spectrometers which were instrumented with three
horizontal drift chambers for the track reconstruction followed by
three scintillator detectors for timing and particle identification.
Three identical OOPS modules were placed symmetrically at azimuthal
angles $\phi^*_{\pi q} = 60^\circ$, $90^\circ$, and $180^\circ$ with
respect to the momentum transfer direction for the measurement at
central kinematics of $\theta^*_{\pi q} = 44.45^\circ$; thus we were
able to isolate the $\sigma_{TT}$, $\sigma_{LT}$, and  $\sigma_{0}$
partial cross sections. An OOPS spectrometer was positioned along
the momentum transfer direction thus directly measuring the parallel
cross section at $\theta^*_{\pi q} = 0^\circ$. A high duty factor
950 MeV electron beam of 7~$\mu$A average current was scattered from
a cryogenic liquid-hydrogen target. Measurements were taken at $W =
1232$ MeV and at four-momentum transfer of $Q^2 = 0.127$
(Gev/c)$^2$. Point cross sections were derived from the finite
acceptances by projecting the measured values to point kinematics
using theoretical models \cite{sato,dmt00,kama,mai00}; the
projection to central kinematics had minimal influence on the
systematic uncertainty of the results. Two simultaneous redundant
measurements were performed at $W = 1232$ MeV, $Q^2 = 0.127$
(Gev/c)$^2$ and $\theta^*_{\pi q} = 44.45^\circ$ by placing two OOPS
spectrometers symmetrically with respect to the scattering plane at
$\phi^*_{\pi q} = 60^\circ$ and $-60^\circ$. The results of both
cross sections were in excellent agreement thus confirming our good
understanding of the OOPS system. Radiative corrections were applied
to the data \cite{kirk} using the code \texttt{ExcluRad}
\cite{afan}. The coincidence time-of-flight spectrum and the
reconstructed missing-mass are presented in Fig.~\ref{fig:tof} and
Fig.~\ref{fig:mmiss} respectively. Elastic scattering data for
calibration purposes were taken using liquid-hydrogen and carbon
targets. The uncertainty in the determination of the central
momentum was 0.1\% and 0.15\% for the pion and electron arms,
respectively while the uncertainty and the spread of the beam energy
were 0.3\% and 0.03\%, respectively. A detailed description of all
experimental uncertainties and their resulting effects in the
measured  partial cross sections is presented in \cite{kirk}.

\begin{figure*}[t]
\centerline{\psfig{figure=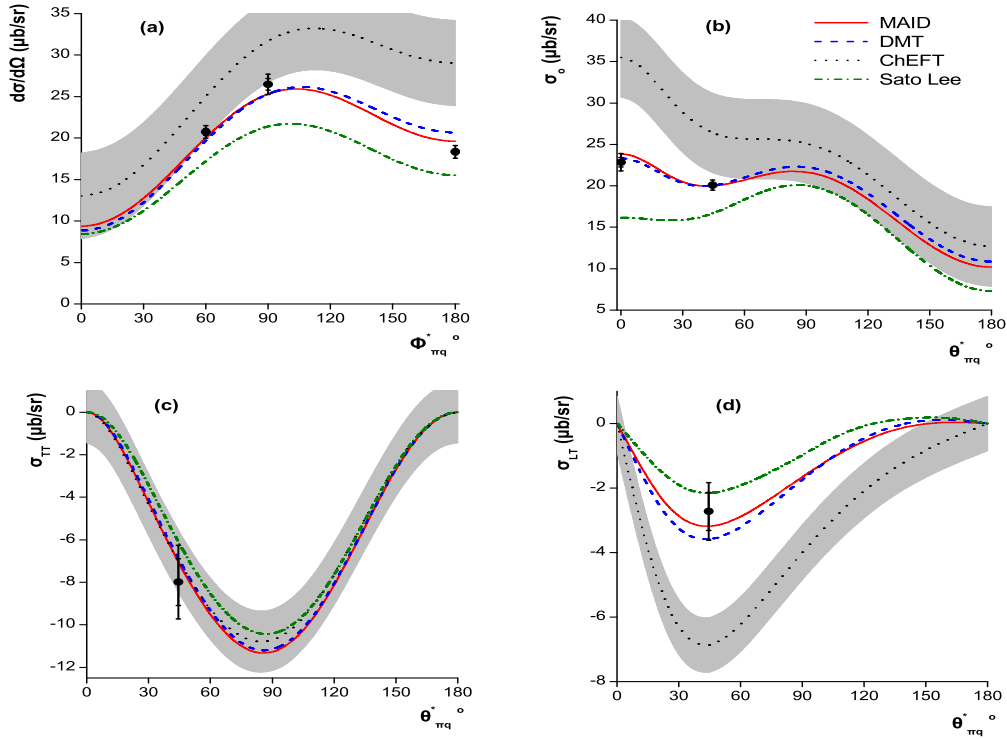,width=15.0cm,height=11.0cm}}
\smallskip
\caption{The experimental results are presented along with the
corresponding theoretical calculations of MAID, DMT, Sato-Lee and
ChEFT. The shaded band corresponds to the uncertainty of the ChEFT
calculation. In (a) the measured cross sections at $\theta^*_{\pi q}
= 44.45^\circ$ are presented. In (b), (c) and (d) the extracted
$\sigma_{0}$, $\sigma_{TT}$ and $\sigma_{LT}$ are presented
respectively.} \label{fig:results}
\end{figure*}

In Fig.~\ref{fig:results} we present the experimental results for
the measured cross sections as well as for the extracted
$\sigma_{0}$, $\sigma_{LT}$ and  $\sigma_{TT}$ partial cross
sections. Both the statistical and the total experimental
uncertainties are exhibited in the figure. The experimental results
are compared with the phenomenological model MAID~2007
\cite{mai00,kama}, the dynamical calculations of Sato-Lee
\cite{sato} and of DMT \cite{dmt00} and the ChEFT calculation of
Pascalutsa and Vanderhaegen \cite{pv}. The MAID model  which offers
a flexible phenomenology and which provides an overall consistent
agreement with the H$(e,e^\prime p)\pi^0$ data at the same $Q^2$
\cite{spaprl} exhibits the best agreement with the experimental
results. The DMT and Sato-Lee are dynamical reaction models which
include pion cloud effects and both calculate the resonant channels
from dynamical equations. DMT uses the background amplitudes of MAID
with some small modifications while Sato-Lee calculate all
amplitudes consistently within the same framework with only three
free parameters. DMT exhibits an overall agreement with the data
with the partial exception of the cross section measurement at
$\theta^*_{\pi q} = 44.45^\circ$, $\phi^*_{\pi q} = 180^\circ$. On
the other hand the Sato-Lee model exhibits a clear disagreement with
the data. The model clearly underestimates the measured cross
sections and the extracted $\sigma_{0}$. For $\theta^*_{\pi q}$
lower than $30^\circ$ one can also point out a qualitative
disagreement of Sato-Lee both with the data and with the rest of the
models; Sato-Lee predicts a plateau below $\theta^*_{\pi q} =
30^\circ$ for $\sigma_{0}$ while the data and the rest of the
theoretical calculations indicate a $\sigma_{0}$ increase as we
approach $\theta^*_{\pi q} = 0^\circ$. On the other hand the
calculation  is within good agreement with the extracted
$\sigma_{LT}$ and $\sigma_{TT}$ cross sections. The magnitude of the
disagreement between the Sato-Lee prediction and the H$(e,e'
\pi^+)n$  results is a bit surprising if we consider the reasonable
agreement of the Sato-Lee calculation with the experimantal results
of the H$(e,e^\prime p)\pi^0$ channel at the same $Q^2$
\cite{spaprl}.

The chiral effective field theory (ChEFT) calculation of Pascalutsa
and Vanderhaegen, is a systematic expansion based on QCD\cite{pv}.
The results of this expansion up to next to leading order is also
presented in Fig.~\ref{fig:results}. Here the contribution of the
next order term in the expansion has been estimated and the
theoretical errors reflect this uncertainty. A significant
overestimation of both the $\sigma_{0}$ and $\sigma_{LT}$ results,
even with the relatively large theoretical uncertainty, indicates
that the next order calculation is required. It is worth pointing
out that  this calculation is in reasonable agreement with the
$\sigma_{0}$ and $\sigma_{LT}$ results for the $p(e,e'p)\pi^0$
channel \cite{stave,spaplb}.

The pion electroproduction database for the $p(e,e'p)\pi^0$ reaction
channel is by now rather extensive resulting in a good understanding
of the resonant amplitudes. Because those measurements are twice as
sensitive to the resonant $I = 3/2$ multipoles as the present
$\pi^+$ measurements, the exploration of the $\pi^+$ channel offers
higher sensitivity to the background amplitudes which can be
observed in the dependence of the cross section as a function of
$\phi^*_{\pi q}$ in Fig.~\ref{fig:results}(a). The resonant $I=3/2$
multipoles dominate the cross section towards $\phi^*_{\pi q} =
0^\circ$ while towards the $\phi^*_{\pi q} = 180^\circ$ region it is
the $I=1/2$ background ones that play the dominant role. One can
observe that the MAID, DMT and Sato-Lee models exhibit an agreement
in their description of the cross section at $\phi^*_{\pi q} =
0^\circ$, where the resonant terms dominate, while they deviate as
we move towards $\phi^*_{\pi q} = 180^\circ$. This is mainly a
consequence of the disagreement of the background amplitudes  of the
models. These conclusions reinforce those that have been inferred
from the background sensitive results previously acquired for the
$p(e,e'p)\pi^0$ channel at  $Q^2$= 0.06 and 0.20 $(GeV/c)^{2}$. at
the lower wing of the resonance $(W=1155$~MeV) \cite{staveprc} which
exhibit high sensitivity to background amplitude contributions. This
study also shows a relatively better description of the background
amplitudes from the MAID and the DMT models compared to the Sato-Lee
one, after the resonant amplitudes have been refitted to the
resonant data, but at the same time none the models  completely
succeeds in agreeing with the data. In contrast to these findings
the  Sato-Lee dynamical model works well for the CLAS experiment for
the $e p \rightarrow e^{'} \pi^{+} n$ reaction at  $0.25 \leq Q^{2}
\leq  0.65~(GeV/c)^{2}$\cite{clas-pip}. In addition the Sato-Lee
model is in good agreement with the $\sigma_{LT^{'}}$ data on
resonance \cite{stave,spaplb} at $Q^{2}$ = 0.06 and 0.20
$(GeV/c)^{2}$ which also provide additional information regarding a
part of the background amplitude contributions.

In conclusion, we present new precise measurements of the
$p(e,e'\pi^+)n$ reaction channel in the $\Delta(1232)$ resonance
region. The new data, which exhibit a higher sensitivity to the
background amplitude terms compared to our previously  explored
$p(e,e'p)\pi^0$ channel studies in the small $Q^{2}$ region
\cite{spaprc,kun00,spaprl,stave,spaplb,staveprc}, provide strong
constraints to the most recent theoretical calculations. We find
that the DMT \cite{dmt00} and MAID \cite{mai00} models are in
reasonable agreement with the data but that the Sato-Lee model
\cite{sato} and the chiral effective field theory calculations
\cite{pv} are not.


\begin{references}
\bibitem{notation}The spin-parity
selection rules in the $N (J^\pi = 1/2^+) \rightarrow \Delta  (J^\pi
= 3/2^+)$ transition allow only the presence of a magnetic dipole
($M1$) and electric quadrupole ($E2$) or Coulomb quadrupole ($C2$)
photon absorption amplitudes, or the corresponding pion production
multipoles $M^{3/2}_{1+}$, $E^{3/2}_{1+}$, and $S^{3/2}_{1+}$ to
contribute. The notation for the latter is\cite{multi} that M, E, S
stand for magnetic, electric, or scalar, the superscript is the
isospin channel = 1/2 or  3/2 (3/2 for the resonant terms), and  the
subscript stands for the orbital angular momentum transfer L, and
the $\pm$ stands for the total angular momentum $J = L \pm 1/2$ (
the 1+ label is for J = 3/2).
\end{references}
\end{document}